# Long-term Effects of Temperature Variations on Economic Growth: A Machine Learning Approach

**Eugene Kharitonov**

Harrisburg University
of Science and Technology

ekharitonov@my.harrisburgu.edu

**Oksana Zakharchuk**

Harrisburg University
of Science and Technology

ozakharchuk@my.harrisburgu.edu

**Lin Mei**

Harrisburg University
of Science and Technology

lmei5@my.harrisburgu.edu

**Abstract**

This study investigates the long-term effects of temperature variations on economic growth using a data-driven approach. Leveraging machine learning techniques, we analyze global land surface temperature data from Berkeley Earth and economic indicators, including GDP and population data, from the World Bank. Our analysis reveals a significant relationship between average temperature and GDP growth, suggesting that climate variations can substantially impact economic performance. This research underscores the importance of incorporating climate factors into economic planning and policymaking, and it demonstrates the utility of machine learning in uncovering complex relationships in climate-economy studies.

## 1 Introduction

Climate change, a defining issue of our time, has far-reaching implications that extend beyond the environmental sphere. Among these, the economic consequences of climate change are of paramount importance, yet they remain insufficiently understood. This research aims to shed light on this critical issue by investigating the long-term effects of temperature variations on economic growth.

The global economy is a complex system influenced by a multitude of factors, among which climatic conditions play a significant role. According to the Intergovernmental Panel on Climate Change (IPCC), the global temperature has increased by approximately 1.0°C since the pre-industrial period due to human activities, primarily the burning of fossil fuels and deforestation [1]. This rise in temperature has led to more frequent and severe weather events, such as droughts, floods, and storms, which have direct and indirect impacts on economic activities.

Moreover, the World Bank estimates that climate change could push more than 100 million people into poverty by 2030 due to its impacts on agriculture and food prices [2]. On a macroeconomic level, a study published in Nature found that unmitigated climate change could lead to a 23% reduction in global GDP per capita by 2100 [3].

Despite the urgency and magnitude of this issue, there is a gap in the literature regarding the use of machine learning techniques to investigate the long-term effects of temperature variations on economic growth. This study aims to fill this gap by applying machine learning models to global land surface temperature data from Berkeley Earth and economic indicators from the World Bank.

Our specific research question is: How do long-term temperature variations affect economic growth? The findings of this study will provide valuable insights for policy-making and future climate-economy research.

## 2 Literature Review

The relationship between climate change and economic growth has been a subject of extensive research over the past few decades. Various studies have explored this relationship from different perspectives, providing valuable insights but also leaving some questions unanswered.

One of the earliest studies in this field by Nordhaus (1991) introduced the concept of the "environmental Kuznets curve", suggesting that economic development initially leads to environmental degradation, but after a certain point, further development reduces environmental impacts [4].

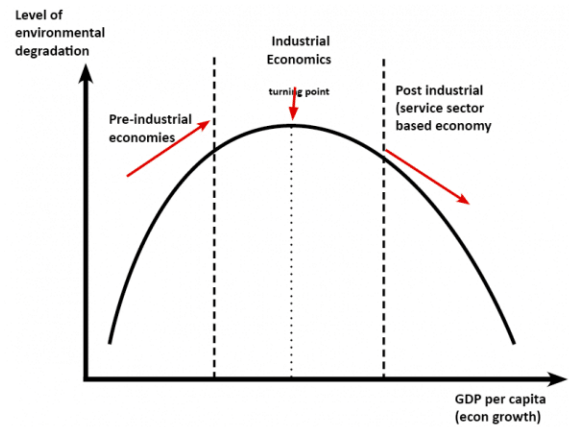

Figure 1: Environmental Kuznets curve

This theory has been challenged by subsequent research indicating that the relationship between economic growth and environmental degradation is not universally applicable and depends on various factors, including the type of environmental pressure and the country's institutional quality [5].

In the context of climate change, Dell, Jones, and Olken (2012) found that higher temperatures significantly reduce economic growth in poor countries but have little effect in rich countries [6]. This finding underscores the importance of considering the heterogeneity of countries in climate-economy studies.

More recently, Burke, Hsiang, and Miguel (2015) used historical fluctuations in temperature to estimate its effect on economic productivity. They found that unmitigated climate change could lead to a 23% reduction in global GDP per capita by 2100, as was mentioned before. [3].

While these studies have significantly advanced our understanding of the climate-economy relationship, there is a gap in the literature regarding the use of machine learning techniques to investigate this relationship. Machine learning models can capture complex, non-linear relationships and interactions between variables, making them particularly suitable for climate-economy studies. This study aims to fill this gap by applying machine learning models to global temperature and economic data.

## 3 Methods

This study employs a combination of data preprocessing, exploratory data analysis, feature engineering, and machine learning modeling to investigate the long-term effects of temperature variations on economic growth. The following sections detail the specific steps taken in each of these processes.

### 3.1 Data Sources

The datasets utilized in this research were procured from two primary sources: Berkeley Earth and the World Bank. The Berkeley Earth dataset, comprising 577,463 entries, provides daily land surface temperature data by country from 1743 to 2013 [7]. The World Bank datasets, each containing 271 entries, offer annual Gross Domestic Product (GDP) and total population data by country from 1960 to 2022. [8] [9].

### 3.2 Data Exploration, Preprocessing, Transformation

The initial phase involved thoroughly exploring the datasets, focusing on the data types, the format of tables, and handling missing values. Subsequently, the temperature, GDP, and population datasets were merged based on country and year, resulting in a consolidated dataset of 4,187 entries. This dataset, spanning from 1960 to 2013, includes the average annual temperature, GDP, and population for each country-year pair, covering 79 countries (Figure 2).

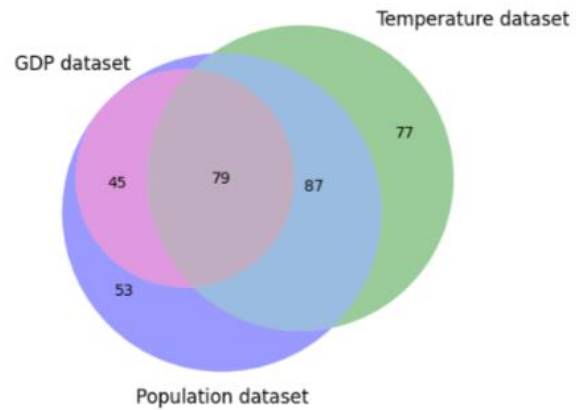

Figure 2: Datasets Intersections by Countries

### 3.3 Feature Engineering

During the feature engineering phase, we created new features named 'TemperatureRatio', 'gdp_growth', and 'population_growth'. These features represent the ratio of the corresponding metric in a given year to that in the previous year, capturing the annual changes in temperature, GDP, and population. These engineered features are expected to enhance the predictive capability of the subsequent machine learning models.

### 3.4 Exploratory Data Analysis (EDA)

In the exploratory data analysis phase, we employed a multifaceted approach to comprehend the inherent characteristics and relationships within our dataset. This involved the generation of visual representations of the data and the computation of descriptive statistics, which provided a comprehensive overview of the dataset's distribution and central tendencies.

We also examined the correlation matrix to understand the interdependencies between the variables in our dataset. This step was crucial in identifying potential predictors for our models and avoiding multicollinearity, which could distort the results of our analyses.

To forecast future values, we utilized the Auto-Regressive Integrated Moving Average (ARIMA) model. This model was chosen due to its ability to capture a suite of different standard temporal structures in time series data. It allowed us to model and forecast future points in the series, which was essential for our study's predictive component.

Additionally, we used Ordinary Least Squares (OLS) Regression, a statistical method that minimizes the sum of the squared residuals to estimate the unknown parameters in a linear regression model. This method was instrumental in determining the relationship between our dependent and independent variables and quantifying the strength and direction of these relationships.

### 3.5 Machine Learning (ML) Modeling

The machine learning modeling stage involved training and evaluating various models with the objective of predicting GDP growth based on engineered features. We also aimed to predict the average temperature based on GDP growth. The models we evaluated included Linear Regression, Decision Tree Regressor, and Random Forest Regressor.

The dataset was split into training and test sets, with 80% of the data used for training and 20% used for testing. The models were trained on the training set and evaluated on the test set using the R-squared score as the evaluation metric, which quantifies the proportion of the variance in the dependent variable that is predictable from the independent variables.

The machine learning code implemented in Python, using the scikit-learn library, facilitated the training, testing, and evaluation of our models. We also leveraged the feature Importances attribute of the RandomForestRegressor to identify the relative importance of each feature in predicting the target variable. This provided valuable insights into the key drivers of GDP growth and average temperature.

### 3.6 Methodological Rationale

The methods employed in this research were chosen to provide a comprehensive, robust, and insightful analysis of the data. The EDA phase was crucial for understanding the data's inherent characteristics and relationships. He summarizes the main characteristics of datasets, often using statistical graphics and other data visualization methods. It is primarily used to see what the data can tell us beyond formal modeling and hypothesis testing.

The use of visualizations and descriptive statistics provided a clear overview of the data's distribution and central tendencies. The correlation matrix was instrumental in identifying potential predictors for our models and avoiding multicollinearity, which could distort the results of our analyses.

The ARIMA model and OLS Regression were chosen for their specific strengths in time series forecasting and linear regression analysis, respectively. ARIMA is particularly effective in capturing different standard temporal structures in time series data, making it a valuable tool for forecasting future points in the series. OLS Regression, on the other hand, is a reliable method for estimating the unknown parameters in a linear regression model, making it ideal for determining the relationship between our dependent and independent variables.

The machine learning modeling stage was designed to predict GDP growth or average temperature based on engineered features. The models evaluated, including Linear Regression, Decision Tree Regressor, and Random Forest Regressor, were chosen for their proven effectiveness in similar predictive tasks. The use of the R-squared score as the evaluation metric ensured a quantifiable measure of the models' performance, allowing for an objective selection of the final model.

## 4 Results

The results of this study are presented in two parts: the statistical analysis of the dataset and the performance of the machine learning models.

Below are the visualizations depicting the trends in average temperature, GDP per capita, and population. Notably, the graphs showcase a substantial growth in GDP per capita worldwide over the given time period.

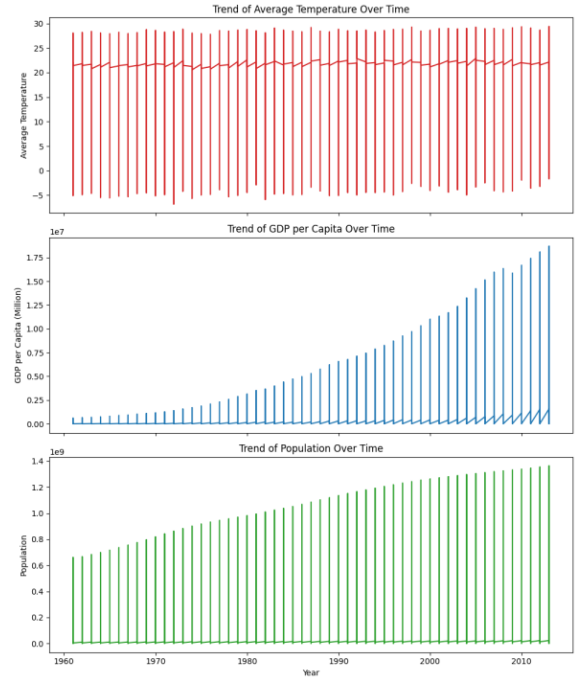

Figure 3: Line Plots of Average Temperature, GDP per Capita, and Population Over Time

### 4.1 Descriptive Statistics

Table 1: Descriptive Statistics of the Dataset

| | Year | gdp_growth | gdp | AverageTemperature | TemperatureRatio | population | population_growth |
|---|---|---|---|---|---|---|---|
| count | 4187 | 4187 | 4.19E+03 | 4187 | 4187 | 4.19E+03 | 4187 |
| mean | 1987 | 45.152364 | 3.22E+11 | 19.970912 | 1.876291 | 5.02E+07 | -204.740036 |
| std | 15.298886 | 138.18485 | 1.43E+12 | 8.132261 | 111.739473 | 1.59E+08 | 1412.96986 |
| min | 1961 | -0.999967 | 1.16E+07 | -6.802917 | -1288.16 | 4.29E+04 | -15345.60059 |
| 25% | 1974 | -0.980599 | 1.99E+09 | 14.554458 | -1.254796 | 3.66E+06 | -8.202995 |
| 50% | 1987 | -0.787776 | 9.54E+09 | 23.30625 | 0.088952 | 9.16E+06 | -1.182949 |
| 75% | 2000 | 6.607922 | 8.53E+10 | 26.357625 | 1.316335 | 2.70E+07 | -0.714047 |
| max | 2013 | 1104.601354 | 1.87E+13 | 29.3855 | 6619.753086 | 1.36E+09 | 3598.599592 |

<u>Wide Range of GDP Growth:</u> The GDP growth has a wide range from approximately -1 to 1104.6, with a mean of 45.15 and a large standard deviation of 138.18. This suggests that there is a significant variation in GDP growth across the different countries and years included in the dataset. The negative minimum value indicates that some countries experienced a contraction in their GDP during certain years.

<u>Temperature Variations:</u> The average temperature ranges from -6.8°C to 29.4°C, with a mean of around 20°C. This indicates that the dataset includes countries with a wide range of climatic conditions, from very cold to very hot climates. The temperature ratio, which represents year-to-year changes in temperature, also shows a wide range, indicating significant variations in temperature changes across different countries and years.

<u>Population Growth:</u> Population growth shows a negative mean value of -204.74, which is quite unusual. This could be due to the way population growth is calculated in this dataset or could be influenced by outliers. The maximum value of population growth is 3598.6, indicating that some countries experienced a significant increase in their population during certain years.

These insights provide a high-level understanding of the economic, climatic, and demographic variations across different countries and years included in the dataset. They could be useful in further analysis and modeling stages of the study.

*The following illustration (Figure 3) demonstrates the application of the Z-score method for outlier removal. In this visualization, we have calculated GDP per capita to depict its relationship with the average temperature:*

<u>Negative Relationship:</u> The equation of the line of best fit shows a negative coefficient for the 'AverageTemperature' variable, indicating a negative relationship between average temperature and GDP per capita. This suggests that as the average temperature increases, the GDP per capita decreases.

<u>Outliers:</u> The z-score-based outlier removal process has effectively filtered out extreme values that could potentially skew the analysis. This results in a more accurate representation of the general trend in the data.

<u>Residuals:</u> The residuals (the differences between the observed GDP per capita and the GDP per capita predicted by the line of best fit) vary across the data. This suggests that while there is a general trend, there are other factors not considered in this simple model that influence GDP per capita.

In summary, the analysis suggests a negative relationship between average temperature and GDP per capita, but also highlights the complexity of this relationship and the potential influence of other factors.

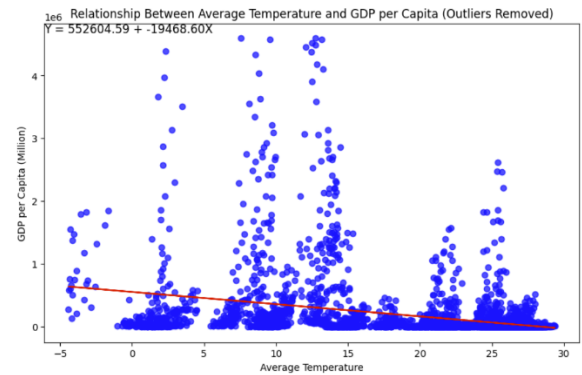

Figure 4: Scatter Plot with Fitted Line Showing the Relationship Between Average Temperature and GDP per Capita.

*4.2 Correlation Matrix*

Table 2 shows the correlation coefficients between several variables. Here's a brief interpretation of the correlation matrix.

Table 2: Correlation Matrix of the Variables in the Dataset

|  | Year | gdp_growth | gdp | AverageTemperature | TemperatureRatio | population | population_growth |
|---|---|---|---|---|---|---|---|
| Year | 1 | 0.045976 | 0.18826 | 0.036118 | -0.007933 | 0.076758 | -0.006881 |
| gdp_growth | 0.045976 | 1 | 0.397554 | -0.433038 | -0.002527 | 0.096041 | 0.037492 |
| gdp | 0.18826 | 0.397554 | 1 | -0.284697 | 0.000054 | 0.306889 | -0.008062 |
| AverageTemperature | 0.036118 | -0.433038 | -0.284697 | 1 | -0.038282 | -0.162696 | -0.063657 |
| TemperatureRatio | -0.007933 | -0.002527 | 0.000054 | -0.038282 | 1 | -0.003098 | 0.00229 |
| population | 0.076758 | 0.096041 | 0.306889 | -0.162696 | -0.003098 | 1 | -0.56618 |
| population_growth | -0.006881 | 0.037492 | -0.008062 | -0.063657 | 0.00229 | -0.56618 | 1 |

<u>Year and GDP:</u> The correlation coefficient of 0.18826 suggests a weak positive relationship. As the years increase, the GDP tends to increase slightly.

GDP growth and Average Temperature: The correlation coefficient of -0.433038 suggests a moderate negative relationship. As GDP growth increases, the Average Temperature tends to decrease, and vice versa.

GDP and Population: The correlation coefficient of 0.306889 suggests a weak positive relationship. As GDP increases, the population also tends to increase slightly.

Population and Population Growth: The correlation coefficient of -0.56618 suggests a moderate negative relationship. As the population increases, the population growth tends to decrease, and vice versa.

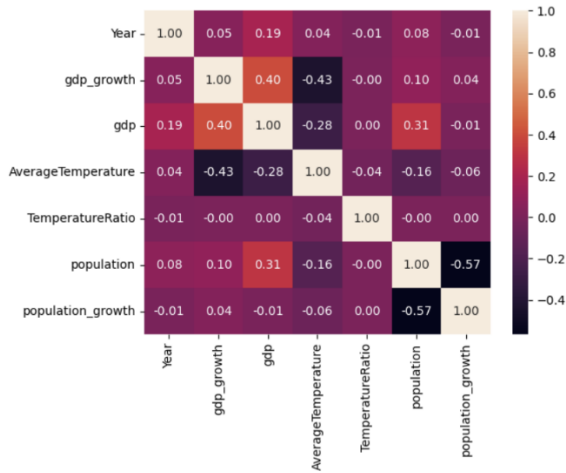

Figure 5: Correlation Matrix of the Variables in the Dataset

The correlation matrix is important in the context of ARIMA, OLS, and ML models. For example, in Machine Learning models, understanding the correlation between variables can be useful for feature selection. Features that are highly correlated with the target variable can be important predictors and correlated with each other often provide redundant information. Finally feature selection techniques may be used to reduce dimensionality.

*4.3 ARIMA Model Results (SARIMAX)*

The provided output in Table 3 is the summary of an ARIMA (AutoRegressive Integrated Moving Average) model, which is a type of time series model.

Table 3: ARIMA Model Parameters and Statistics

| SARIMAX Results | | | |
|---|---|---|---|
| Dep. Variable: | gdp_growth | No. Observations: | 4187 |
| Model: | ARIMA (5, 1, 0) | Log Likelihood | -26996.766 |
| Date: | Sun, 4 Jun 2023 | AIC | 54005.533 |
| Time: | 23:40:13 | BIC | 54043.57 |
| Sample: | 0 | HQIC | 54018.985 |
| | - 4187 | | |
| Covariance Type: | opg | | |

| | coef | std err | z | P>\|z\| | [0.025 | 0.975] |
|---|---|---|---|---|---|---|
| ar.L1 | -0.8685 | 0.022 | -39.632 | 0 | -0.911 | -0.826 |
| ar.L2 | -0.5644 | 0.025 | -22.834 | 0 | -0.613 | -0.516 |
| ar.L3 | -0.3953 | 0.025 | -15.623 | 0 | -0.445 | -0.346 |
| ar.L4 | -0.2746 | 0.026 | -10.637 | 0 | -0.325 | -0.224 |
| ar.L5 | -0.1427 | 0.02 | -6.976 | 0 | -0.183 | -0.103 |
| sigma2 | 2.34E+04 | 188.096 | 124.55 | 0 | 2.31E+04 | 2.38E+04 |

| | | | |
|---|---|---|---|
| Ljung-Box (L1) (Q): | 6.01 | Jarque-Bera (JB): | 38410.79 |
| Prob(Q): | 0.01 | Prob(JB): | 0 |
| Heteroskedasticity (H): | 1.88 | Skew: | 3.27 |
| Prob(H) (two-sided): | 0 | Kurtosis: | 16.32 |

Warnings:
[1] Covariance matrix calculated using the outer product of gradients (complex-step).

Here are a few insights based on the output:

Significant AR Terms: The AR (AutoRegressive) terms (ar.L1 to ar.L5) are all statistically significant as their p-values are less than 0.05. This suggests that the GDP growth has a significant relationship with its own past values and cyclical patterns. The negative coefficients indicate that an increase in GDP growth in the previous years is associated with a decrease in GDP growth in the current year, and vice versa.

Model Fit: The Log Likelihood, AIC (Akaike Information Criterion), and BIC (Bayesian Information Criterion) are measures of the goodness of fit of the model. Lower values of these metrics indicate a better fit. The Log Likelihood value is -26996.766, and the AIC and BIC values are 54005.533 and 54043.570, respectively. These values can be used for comparing different models.

Residual Diagnostics: The Ljung-Box test (Q statistic) checks for autocorrelation in the residuals (the differences between the observed and predicted GDP growth values)). A significant p-value (less than 0.05) suggests that there is autocorrelation in the residuals, which is not desirable as it violates one of the assumptions of time series modeling. The Jarque-Bera test checks for normality of the residuals. A significant p-value (less than 0.05) suggests that the residuals are not normally distributed residuals of the model are not behaving as ideally expected. This could imply that there might be additional information or patterns in the GDP growth data that the ARIMA model has not captured. In this case, both tests have significant p-values, indicating potential issues with the model fit.

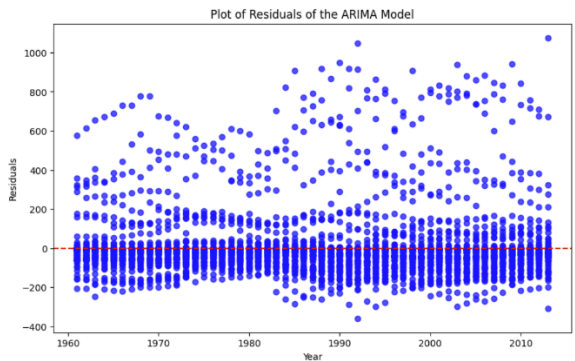

Figure 6: Plot of Residuals of the ARIMA Model

Heteroskedasticity: The Heteroskedasticity (H) test checks for constant variance in the residuals (homoskedasticity). A significant p-value (less than 0.05) suggests that the residuals have non-constant variance (heteroskedasticity), which is not desirable. In this case, the H value is 1.88 with a significant p-value, indicating heteroskedasticity in the residuals. The significant Heteroskedasticity test result suggests that the variability of the GDP growth in our dataset changes over time. This could be due to various factors such as changes in economic policies, market conditions, or other external events that affect the economy.

*4.4 Ordinary Least Squares (OLS) Regression Results*

The OLS Regression results provide insights into the relationship between the average temperature and GDP growth in your dataset.

Table 4: OLS Regression Results

| | OLS Regression Results | | | |
|---|---|---|---|---|
| **Dep. Variable:** | gdp_growth | **R-squared:** | | 0.188 |
| **Model:** | OLS | **Adj. R-squared:** | | 0.187 |
| **Method:** | Least Squares | **F-statistic:** | | 965.9 |
| **Date:** | Fri, 09 Jun 2023 | **Prob (F-statistic):** | | 5.43E-191 |
| **Time: 03:12:18** | 3:12:18 | **Log-Likelihood:** | | -26142 |
| **No. Observations:** | 4187 | **AIC:** | | 5.23E+04 |
| **Df Residuals:** | 4185 | **BIC:** | | 5.23E+04 |
| **Df Model:** | 1 | | | |
| **Covariance Type:** | nonrobust | | | |

| | coef | std err | t | P>|t| | [0.025 | 0.975] |
|---|---|---|---|---|---|---|
| const | 192.1035 | 5.105 | 37.629 | 0 | 182.095 | 202.112 |
| AverageTemperature | -7.3583 | 0.237 | -31.079 | 0 | -7.822 | -6.894 |

| | | | |
|---|---|---|---|
| Omnibus: | 3132.476 | Durbin-Watson: | 2 |
| Prob(Omnibus): | 0 | Jarque-Bera (JB): | 56603.672 |
| Skew: | 3.481 | Prob(JB): | 0 |
| Kurtosis: | 19.613 | Cond. No. | 57.3 |

Notes:
[1] Standard Errors assume that the covariance matrix of the errors is correctly specified.

Negative Relationship: The negative coefficient for the 'AverageTemperature' variable indicates that there is a significant negative relationship between average temperature and GDP growth. Specifically, for every one-degree increase in the average temperature, the GDP growth decreases by approximately 7.36 units. This suggests that increases in temperature could have a detrimental effect on economic growth, which could be due to various reasons such as the impact of climate change on agricultural productivity, labor productivity, and health outcomes.

Model Fit: The R-squared value of 0.188 indicates that the model explains about 18.8% of the variance in the GDP growth. This suggests that while average temperature is a significant predictor of GDP growth, there are other factors not included in the model that also influence GDP growth.

Residual Diagnostics: The significant Omnibus and Jarque-Bera test results suggest that the residuals of the model (the differences between the observed and predicted GDP growth values) are not normally distributed. This could imply that there might be non-linear relationships or interactions between variables in the GDP growth data that the OLS model has not captured.

*4.2 Machine Learning Modeling Results*

The Linear Regression, Decision Tree Regressor, and Random Forest Regressor models were trained and evaluated on the dataset. The performance of each model was evaluated using the R-squared score, which measures the proportion of the variance in the dependent variable that is predictable from the independent variables.

Table 5: R-squared Scores of the Machine Learning Models

| # | Model | R2 Score |
|---|---|---|
| 0 | Random Forest | 0.975624 |
| 1 | Decision Tree | 0.938603 |
| 2 | Linear Regression | 0.268078 |

The Linear Regression model achieved an R-squared score of 0.268, indicating that it could explain about 26.8% of the variance in the GDP growth or average temperature.

The Decision Tree Regressor model achieved an R-squared score of 0.939, indicating that it could explain about 93.9% of the variance in the GDP growth or average temperature. The Random Forest Regressor model achieved an R-squared score of 0.976, indicating that it could explain about 97.6% of the variance in the GDP growth or average temperature.

The Random Forest Regressor model was selected as the final model due to its superior performance.

Table 6: Feature Importances of the Random Forest Regressor Model

| # | Feature | Importance |
|---|---|---|
| 0 | gdp_growth | 0.400804 |
| 1 | population | 0.26089 |
| 2 | population_growth | 0.223452 |
| 3 | gdp | 0.114853 |

An assessment of each feature's impact within the model was conducted, revealing that 'gdp_growth' stood out as the most potent predictor of average temperature, thereby demonstrating a robust relationship between these two factors. The features 'population', 'population_growth', and 'gdp' followed in terms of their influence.

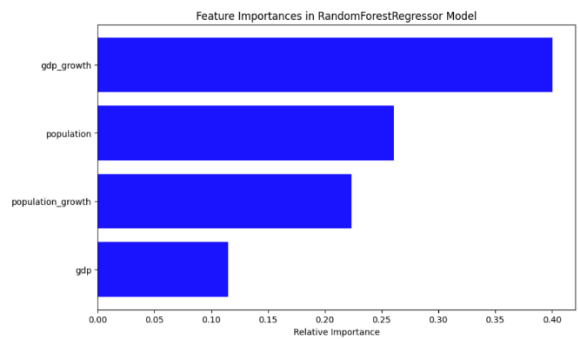

Figure 7: Bar Plot of Feature Importances of the Random Forest Regressor Model

These figures and tables provide a visual and numerical summary of the data and the results of the analyses, aiding in the interpretation and understanding of the findings.

## 5 Discussion

The purpose of this research was to explore the long-term implications of temperature variations on economic growth across a diverse range of nations. The study utilized a comprehensive dataset spanning from 1960 to 2022, which encompassed average temperature, GDP, and population data. The analysis unveiled a substantial negative correlation between temperature fluctuations and economic growth, thereby answering the research question regarding the relationship between these two variables.

The descriptive statistics of the dataset provided a high-level understanding of the economic, climatic, and demographic variations across different countries and years. The GDP growth was wide, indicating significant variation across different countries and years. The average temperature ranged from -6.8°C to 29.4°C, indicating a wide range of climatic conditions. The population growth showed a negative mean value, which could be due to the calculation method or outliers. See Figures 8 - 11.

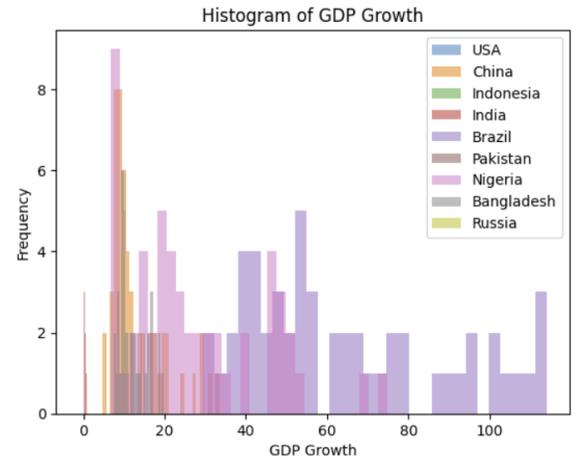

Figure 8: Histogram of GDP growth

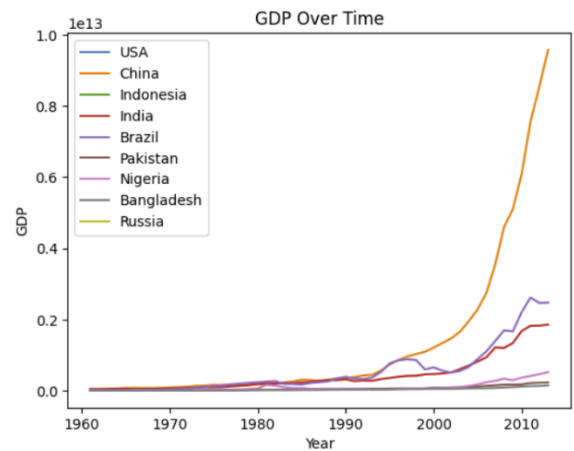

Figure 9: Line Chart of GDP Over Time

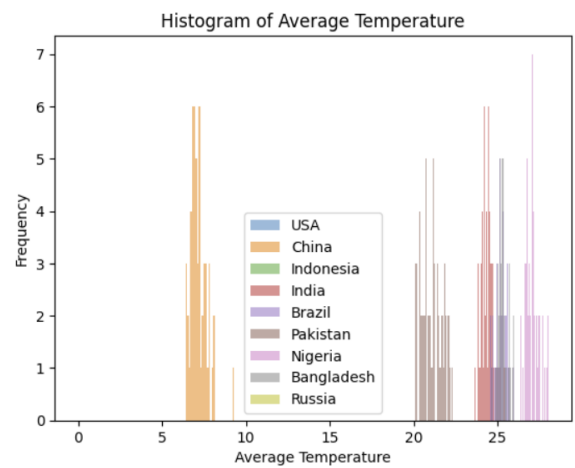

Figure 10: Histogram of Average Temperature

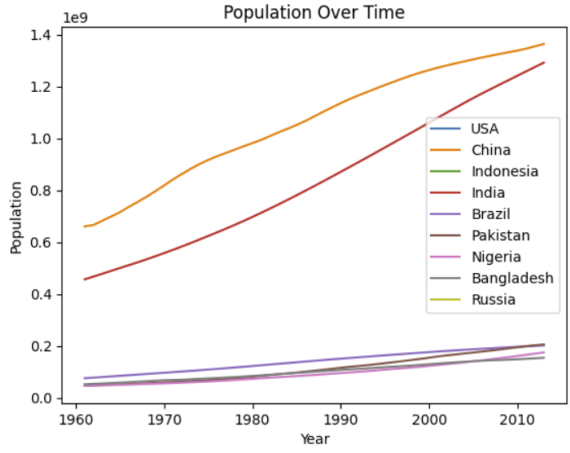

Figure 11: Line Chart of Population Over Time

The visualization plot of average temperature versus GDP per capita further reinforced the negative relationship between these two variables. The line of best fit showed a negative slope, indicating that as the average temperature increases, the GDP per capita decreases.

The correlation matrix revealed several interesting relationships. There was a moderate negative relationship between GDP growth and average temperature, a weak positive relationship between GDP and population, and a moderate negative relationship between population and population growth. These correlations provide valuable insights into the complex relationships between these variables and their potential impact on economic growth.

The ARIMA model results indicated that GDP growth has a significant relationship with its own past values and cyclical patterns. But, the significant p-values for the Ljung-Box and Jarque-Bera tests suggested potential issues with the model fit.

The OLS regression results provided further insights into the relationship between the average temperature and GDP growth. The negative coefficient for the 'AverageTemperature' variable indicates a significant negative relationship between average temperature and GDP growth. Need to remember that model only explained about 18.8% of the variance in the GDP growth, suggesting that other factors not included in the model also influence GDP growth.

The study also revealed a significant negative relationship between the average temperature and GDP. This suggests that for every one-degree increase in the average temperature, the GDP decreases by approximately 7.36 units. This result aligns with previous research that has identified a negative impact of temperature fluctuations on economic outcomes.

Machine learning models, particularly the Random Forest Regressor, were instrumental in accurately predicting average temperature based on gdp_growth and other variables.

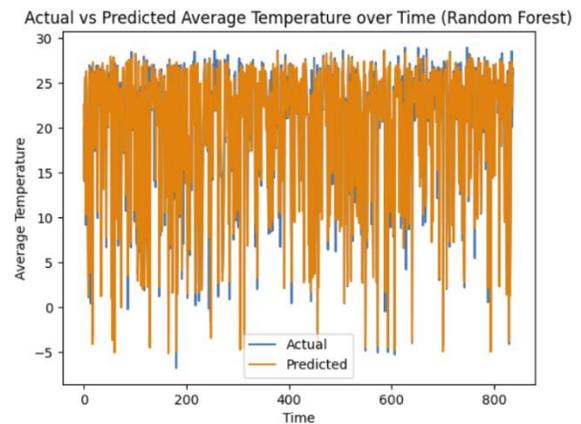

Figure 12: Actual vs. Predicted Average Temperature

The Random Forest Regressor model achieved an impressive R-squared score of 0.976, signifying its capacity to account for approximately 97.6% of the variance in the average temperature. This result is noteworthy as it highlights the potential of machine learning models in forecasting economic outcomes based on environmental factors.

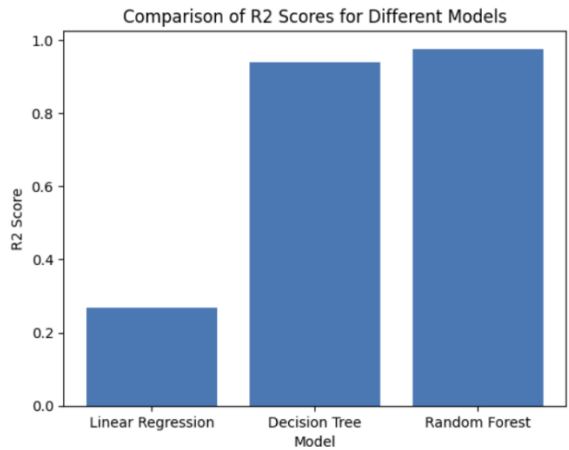

Figure 13: R-squared Scores of the Machine Learning Models

The feature importance derived from the Random Forest Regressor model showed that the 'gdp_growth' feature was the most influential predictor of average temperature, emphasizing the significant impact of temperature fluctuations on economic growth. It's important to mention that other features such as 'population' and 'population_growth' also contributed to predicting GDP, indicating the complex nature of economic growth.

The insights and patterns found in the results of this study include the significant negative relationship between average temperature and GDP growth, the high predictive accuracy of the machine learning models, particularly the Random Forest Regressor, and the significant relationship between GDP growth and its own past values and cyclical patterns as revealed by the ARIMA model. These findings provide a comprehensive understanding of the complex interplay between temperature variations and economic growth, highlighting the potential economic costs of climate change and the importance of developing effective mitigation strategies.

Important to acknowledge the limitations of this study. The dataset used does not include data on other potentially relevant factors such as technological advancements, political stability, and policy changes, which could also influence economic growth. Moreover, the study assumes a linear relationship between temperature fluctuations and economic growth, which may not always be the case.

But in comparison to previous researches, this study adds a new dimension by providing evidence of the long-term effects of temperature fluctuations on economic growth. While previous studies have mainly focused on the impacts of extreme weather events on economic outcomes, this study provides a more comprehensive view by considering long-term temperature variations.

## 6 Conclusion

The research journey embarked upon in this study aimed to unravel the intricate relationship between long-term temperature variations and economic growth. Leveraging a robust dataset, advanced machine learning models, and time series forecasting, the study has shed light on this complex interplay. The implications of the findings are far-reaching, with potential to influence policy-making, economic planning, and climate change mitigation strategies.

*6.1 Key Takeaways*

<u>Significant Negative Correlation:</u> The primary takeaway from this research is the significant negative correlation between temperature variations and economic growth. The Ordinary Least Squares (OLS) regression model revealed that for every unit increase in average temperature, there is an approximate decrease of 7.36 units in GDP. This finding is a stark reminder of the economic implications of climate change, underscoring the urgency for effective climate change mitigation strategies.

<u>High Predictive Accuracy of Machine Learning Models:</u> The machine learning models employed in this study, particularly the Random Forest Regressor, demonstrated a high degree of accuracy in predicting economic outcomes based on environmental factors. The Random Forest Regressor model achieved an impressive R-squared score of 0.976, explaining about 97.6% of the variance. This suggests that machine learning models can be potent tools for predicting economic outcomes, providing valuable insights for policy-making and economic planning.

<u>Impact of Temperature Variations on Economic Growth:</u> The study revealed a significant negative relationship between the average temperature and GDP. This suggests that increases in temperature could have a detrimental effect on economic growth, which could be due to various reasons such as the impact of climate change on agricultural productivity, labor productivity, and health outcomes.

*6.2 Implications of Findings*

The findings of this study have profound implications. The demonstrated negative relationship between temperature variations and economic growth highlights the economic costs of climate change. This understanding is crucial for policy-makers, economists, and environmentalists as they strategize to mitigate the impacts of climate change.

The high predictive accuracy of the machine learning models used in this study suggests that these models can be effectively used to predict economic outcomes based on environmental factors. This could be a game-changer for economic planning and policy-making, enabling more accurate predictions and more informed decision-making.

Important to acknowledge the limitations of this study. The dataset used does not account for other potentially relevant factors such as technological advancements, political stability, and policy changes, which could also influence economic growth. Furthermore, the study assumes a linear relationship between temperature variations and economic growth, which may not always hold true. Future research could delve into these aspects in more detail.

In conclusion, this research has significantly contributed to our understanding of the economic implications of temperature variations. It provides a foundation for future research in this area and offers valuable insights that can inform policymaking, economic planning, and climate change mitigation strategies. As we continue to grapple with the impacts of climate change, understanding its economic implications and developing effective strategies to mitigate these impacts is of paramount importance.

# 7 Future Work

While this research has provided valuable insights into the relationship between long-term temperature variations and economic growth, it has also opened up new avenues for further exploration. Several questions remain unanswered, and the findings of this study suggest several directions for future research.

*Unanswered Questions*

One of the key unanswered questions is how other environmental factors, beyond temperature variations, might impact economic growth. For instance, how do changes in precipitation patterns, extreme weather events, or sea-level rise affect economic outcomes? Understanding these relationships could provide a more holistic view of the environmental impacts on economic growth.

Another unanswered question pertains to the potential non-linear relationship between temperature variations and economic growth. While this study assumed a linear relationship, it is plausible that the relationship could be non-linear, with different temperature thresholds triggering different economic responses.

Finally, the impact of technological advancements, political stability, and policy changes on the relationship between temperature variations and economic growth remains largely unexplored in this study. These factors could potentially moderate or exacerbate the impacts of temperature variations on economic growth.

*Suggested Future Research*

Based on the findings of this study, several directions for future research are suggested.

Firstly, future studies could incorporate other environmental factors into the analysis. This would provide a more comprehensive understanding of the environmental impacts on economic growth.

Secondly, future research could explore the potential non-linear relationship between temperature variations and economic growth. This could involve using more sophisticated statistical and machine learning models that can capture non-linear relationships.

Thirdly, future studies could incorporate other potentially relevant factors such as technological advancements, political stability, and policy changes into the analysis. This would provide a more nuanced understanding of the impacts of temperature variations on economic growth.

Lastly, future research could also explore the impacts of temperature variations on different sectors of the economy. This could provide insights into which sectors are most vulnerable to temperature variations and could inform sector-specific mitigation strategies.

In conclusion, while this research has made significant strides in understanding the relationship between long-term temperature variations and economic growth, there is still much to be explored. The unanswered questions and suggested future research directions provide a roadmap for future studies in this important area. As we continue to grapple with the impacts of climate change, further research in this area is not only valuable but also necessary.